\documentclass[12pt,a4paper]{article}
\usepackage{authblk}
\usepackage{cite}
\usepackage{graphicx}
\usepackage[breaklinks=true, colorlinks=true, linkcolor=blue, 
urlcolor=blue, citecolor=blue]{hyperref}
\newcommand{\PRL}{Phys. Rev. Lett. }
\newcommand{\PR}{Phys. Rev. }

\begin{document}

\title{Numerical and experimental study of the effects of noise on the 
permutation entropy}
\author{C. Quintero-Quiroz\footnote{ 
\url{carlos.alberto.quintero@upc.edu}}}
\author{Simone Pigolotti}
\author{M. C. Torrent}
\author{Cristina Masoller}
\affil{Departament de F\'isica i Enginyeria Nuclear, Universitat 
Polit\`ecnica de Catalunya, Colom 11, Terrassa, 08222 Barcelona, Spain.}
\maketitle

\begin{abstract}
  {We analyze the effects of noise on the permutation entropy of 
  dynamical systems. We take as numerical examples the logistic map 
  and the R\"ossler system. Upon varying the noise strengthfaster, we 
find a 
  transition from an almost-deterministic regime, where the permutation 
  entropy grows slower than linearly with the pattern dimension, to a 
  noise-dominated regime, where the permutation entropy grows faster than 
  linearly with the pattern dimension. We perform the same analysis on 
  experimental time-series by considering the stochastic spiking output 
  of a semiconductor laser with optical feedback. Because of the 
  experimental conditions, the dynamics is found to be always in the 
  noise-dominated regime. Nevertheless, the analysis allows to detect 
  regularities of the underlying dynamics. By comparing the results of 
  these three different examples, we discuss the possibility of 
  determining from a time series whether the underlying dynamics is 
  dominated by noise or not.}
\end{abstract}
\vspace{2pc}

\section{Introduction}

Ordinal analysis is a method of time series analysis that consists of 
computing the probabilities of \emph{ordinal patterns}, which are defined 
according to the ordering of $D$ consecutive values in the series 
\cite{bp2002}. The entropy of these probabilities, referred to as {\em 
permutation entropy}, is a tool to detect possible regularities in the 
time series. In recent years, ordinal patterns and permutation entropy 
have been widely used to investigate complex dynamical systems 
\cite{zanin, special_issue}. They have been employed in the attempt to 
distinguish noise from chaos \cite{amigo2006order, rosso_prl_2007, 
amigo2008combinatorial, aragoneses_2013}, to detect noise-induced order 
\cite{rosso_pre_2009}, serial correlations \cite{tiana_pra_2010} and 
dependencies between two or more time series \cite{groth_pre_2005, 
canovas2011using, bahraminasab2008direction, matilla2011spatial, 
matilla2008non, saco_physicaa_2010, barreiro_chaos_2011}, among many 
other examples. Applications to experimental time series analysis include 
classification and discrimination of dynamical states in normal and 
epileptic EEG \cite{Nicolett-2012, veisi2007fast, li2007predictability, 
bruzzo2008permutation} and detection of heart rate variability under 
different physiological and pathological conditions 
\cite{parlitz2012classifying, frank2006permutation, berg2010comparison}.

Given this growing interest, it is relevant to understand the relation 
between the permutation entropy and other complexity measures. In 
particular, a well-established way to characterize the production of 
information of a dynamical system is the Kolmogorov-Sinai entropy 
$h_{ks}$, see e.g. \cite{beck,compl}. To compute $h_{ks}$, the time 
series is discretized by partitioning the phase space into regions and 
assigning a symbol to each region. Then one computes the probabilities of 
{\em blocks}, which are vectors of $D$ consecutive symbols (more 
details in the following sections). The entropy of the block 
probabilities is the {\em block entropy}. The Kolmogorov-Sinai entropy is 
finally obtained as the rate of growth, for $D\rightarrow \infty$ and in 
the limit of a very refined partition, of the block entropy.

Similarly to $h_{ks}$, one can introduce a {\em permutation entropy rate} 
as the rate of growth for $D\rightarrow \infty$ of the permutation 
entropy. Both the permutation entropy and the Kolmogorov-Sinai entropy 
measure the ``asymptotic'' information rate of representations of the 
time series, the former with ordinal patterns (based in the relative 
order of consecutive values) and the latter with blocks (based on a 
partition of the phase space). The permutation entropy rate and $h_{ks}$ 
are not only conceptually related: for piecewise monotone interval maps 
on the real line, they were shown to be equal \cite{KBP_2002}. This 
result has been later extended to a broad class of dynamics 
\cite{amigo_2005, amigo_2012}. This equivalence is non-trivial 
considering, for example, that the number of total possible ordinal 
patterns grows with $D$ as $D!$, while the number of blocks grows as 
$Q^D$ where $Q$ is the total number of symbols. The two quantities can 
be equal only thanks to the large number of forbidden ordinal patterns, 
strongly limiting the growth of the permutation entropy as $D$ is 
increased.
 
These mathematical results clarify that, under general hypotheses, 
permutation and block entropies share the same asymptotic behavior. 
However, due to difficulties in reaching the asymptotic regime, this 
equivalence can be of little use in many practical cases. For example, 
it has been noted \cite{KBP_2002} that the rate of convergence of the 
permutation entropy to the Kolmogorov-Sinai entropy is extremely slow 
even for one-dimensional maps, while on the contrary, block entropies 
converge very quickly, see e.g. \cite{beck}.

Comparing the two analyses becomes even more problematic for 
high-dimensional and/or noisy dynamics, such as typical experimental 
time-series. Consider for example the extreme case of a time 
series dominated by noise, in which all symbols are equally probable and 
temporal correlations are absent. In this case, the block entropy of 
length $D$ is equal to $D\ln(Q)$, where $Q$ is the total number of 
symbols, while the permutation entropy with patterns of length $D$ is 
equal to $\ln(D!)\sim D\ln D$. This means that the block entropy is 
linear in $D$, with a slope $\ln(Q)$, explicitly dependent on the chosen 
partition, which diverges only in the limit of a very refined partition, 
$Q\rightarrow \infty$. In contrast, the permutation entropy grows more 
than linearly, so that their asymptotic slope is infinite. In both cases, 
the result is an infinite entropy rate. However, to discover it, in the 
first case one needs to construct a very refined partition. In the second 
case, one needs to reach large values of $D$ to appreciate that the slope 
increases logarithmically. Both these tasks can be very difficult when 
analyzing a finite time series due to statistical limitations.

Our goal is to get a better understanding of how noise influences the 
permutation entropy. To this aim, we analyze simulated and experimental 
time series. We mostly focus on permutation entropy as the effect of 
noise on block entropies is fairly well understood, see e.g. 
\cite{cencini_2000, falcioni_2003}. We first analyze time series 
generated from the logistic map and from Poincar\'e sections of the 
three-dimensional R\"ossler system. We conclude with an experimental 
example of output intensity data recorded from a semiconductor laser 
with optical feedback.

\section{Methods}
\subsection{Numerical data}

We consider two dynamical systems: the one-dimensional logistic map and 
the three-dimensional R\"ossler system. In both cases, we study the 
effect of adding to the dynamical equations a Gaussian white noise, 
$\xi_t $ with $\langle\xi_t\rangle=0$ and temporal correlation 
$\langle\xi_t \xi_{t'}\rangle=\delta_{t,t'}$. We considered also the case 
of observational noise (not shown), where the dynamics is deterministic 
but the noise affects the observation, obtaining very similar results.

\subsubsection{Logistic map}
\begin{equation}
 x_{t+1} = 4 x_{t} (1 - x_{t}) + \alpha  \xi_t,
 \label{log}
\end{equation}
where $x_t$ is the state of the system at iteration $t$ and $\alpha$ is 
the noise strength. In order to constrain the variable $x_t$ in the 
interval $[0,1]$, the values of $\xi_t$ that would lead to $x_{t + 1}>1$ 
or $x_{t+1}<0$ are simply discarded and redrawn. Thus, the noise $\xi_t$ 
is temporally uncorrelated, but not purely Gaussian due to this 
truncation effect. To investigate the variation of the permutation 
entropy with the noise strength, we computed the permutation entropy, for 
each value of $\alpha$, from time series of length $N = 12 \times 
10^{7}$. We have also studied other nonlinear one-dimensional maps (Tent, 
Bernoulli and Quadratic) and obtained very similar results to those of 
the logistic map (results not shown).

\subsubsection{R\"ossler system}
The R\"ossler equations read
\begin{eqnarray}
  \dot{X} &=& -  Y  -  Z + \alpha \xi(t) , \nonumber\\
  \dot{Y} &=& X  +  aY \label{eq1}\\
  \dot{Z} &=& b+Z(X-c)\nonumber
\end{eqnarray}
where $\{X,Y,Z\}$ are the states of the system at time $t$, $\alpha$ is 
the noise strength and  $\{a,b,c\}$ are the local parameters set at 
\{$0.1, \,0.1, \,18.0$\}, respectively. 

In order to apply the symbolic methods (ordinal patterns or blocks) we 
need to discretize the dynamics. Instead of employing temporal sampling 
\cite{demicco}, we introduce a Poincar\'e section \cite{review} at $ X = 
0 $, and analyze the time intervals between consecutive crossings of the 
Poincar\'e plane. For each value of $\alpha$, the permutation entropy is 
computed from time-series of $N=12\times10^{7}$ data points.

\subsection{Experimental data}

Experimental data was recorded from the output intensity of a 
semiconductor laser with optical feedback operating in the low-frequency 
fluctuations (LFFs) regime. In this regime, the laser intensity displays 
sudden and apparently random dropouts, followed by gradual recoveries. 
This spiking dynamics has received considerable attention because the 
intensity dropouts are induced by stochastic effects and deterministic 
nonlinearities. The optical feedback introduces a delay which renders 
the system in principle infinite dimensional. Therefore, the laser in the 
LFF regime generates complex fluctuations that, because of the stochastic 
and high-dimensional nature of the underlying dynamics, are suitable to 
be investigated by means of complexity measures such as the permutation 
and block entropies.

The experimental setup is the same as in 
\cite{taciano_optics_express_2015} and uses a 650 nm AlGaInP 
semiconductor laser (SONY SLD1137VS) with optical feedback. The feedback 
was given through a mirror placed 70 cm apart from the laser cavity, 
with a round trip of 4.7 ns. The feedback was controlled using a neutral 
density filter that can adjusts the light intensity injected into the 
laser. The laser has a solitary threshold current of $I_{th} = 28.4$ mA. 
The temperature and current of the laser were stabilized using a combi 
controller Thorlabs ITC501 with an accuracy of $0.01$ C and $0.01$ mA, 
respectively. The current used during the experiment was $I = 29.3$ mA 
and the temperature was set at $T = 17$ C. The neutral density filter was 
adjusted so that the threshold reduction due to feedback was about 7\%. 
The signal was captured using a photo detector (Thorlabs DET210) 
connected to a FEMTO HSA-Y-2-40 amplifier and registered with a 1 GHz 
digital oscilloscope (Agilent Infiniium DSO9104A) with $0.2$ ns of 
sampling. The intensity time series were acquired from the oscilloscope 
by a LabVIEW program that uses a threshold to detect the times when the 
intensity drops, and calculates the time intervals between successive 
threshold crossings (in the following, referred to as  
inter-dropout-intervals, IDIs). We recorded in this way time series of 
more than $10^5$ consecutive IDIs.

\subsection{Methods of analysis}

We compare two different methods to transform a time-series, $x(t)= 
\{x(1)$, $ x(2)\dots x(N)\}$, into a sequence of symbols, $s(t)$: ordinal 
patterns and blocks.

In both cases, one needs to choose a {\em dimension} $D$ for defining 
vectors made up of consecutive entries of the time series, i.e. $\{x(i), 
x(i+1), \dots, x(i+D-1)\}$. Ordinal patterns and blocks differ by the 
way in which the entries of these vectors are transformed into symbols. 
Ordinal patterns classify them according to the ranking (from the largest 
to the smallest value) of the $D$ entries in the vectors. The total 
number of ordinal patterns of length $D$ is then equal to the number of 
permutations, $D!$. For example, with $D = 2$ there are two ordinal 
patterns: $x(t_i) > x(t_{i+1})$ corresponding to the ordinal pattern 
`$01$' and $x(t_i) < x(t_{i+1})$ corresponding to the ordinal pattern 
`$10$'.

For blocks, the phase space is first divided into $Q$ regions, 
associating a symbol to each region. Blocks represent all vectors in 
which each value of the time series correspond to the same symbol. For 
example, let us consider the time series $x (t) = \{0.1,0.6,0.7,0.3\}$, 
and partition the phase space into the two regions $[0, 0.5)$ and $[0.5, 
1]$, associating to them the symbols $0$ and $1$ respectively. With 
$D=2$, the blocks associated to the time series are $\{01,11,10\}$.

Then, the permutation entropy \cite{bp2002} or the block entropy 
\cite{beck, compl} are simply the entropy of the frequency $p_i$ of the 
different patterns in the time series
\begin{equation}
H_D = -\sum_i^M p_i\ln p_i,
\end{equation}
where $M$ is the number of possible patterns: for permutation entropy, 
$M=D!$, while for block entropy, $M=Q^D$.

\section{Results}
\begin{figure}[tp]
\centering
\includegraphics[width=0.495\textwidth]{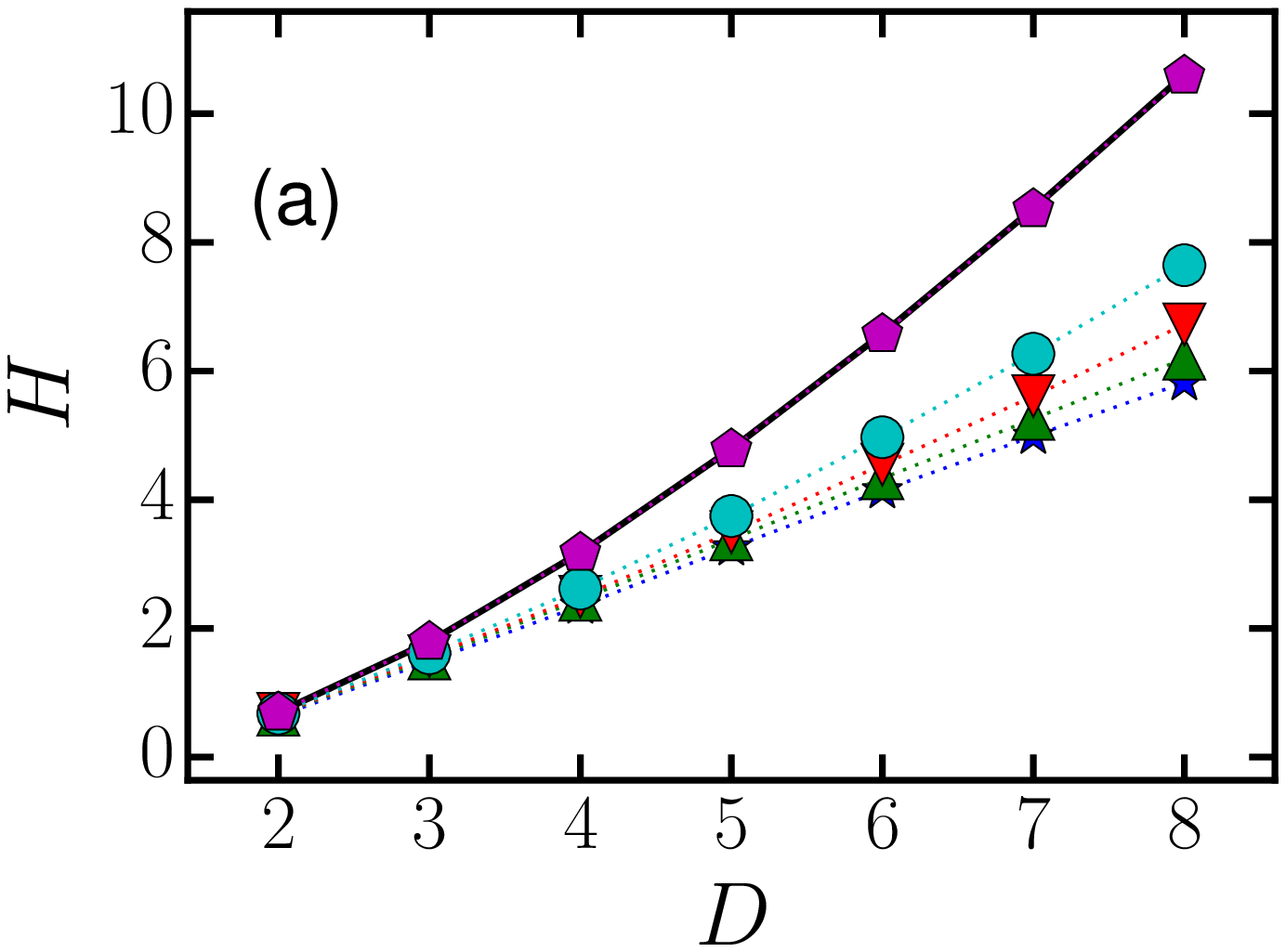}
\includegraphics[width=0.495\textwidth]{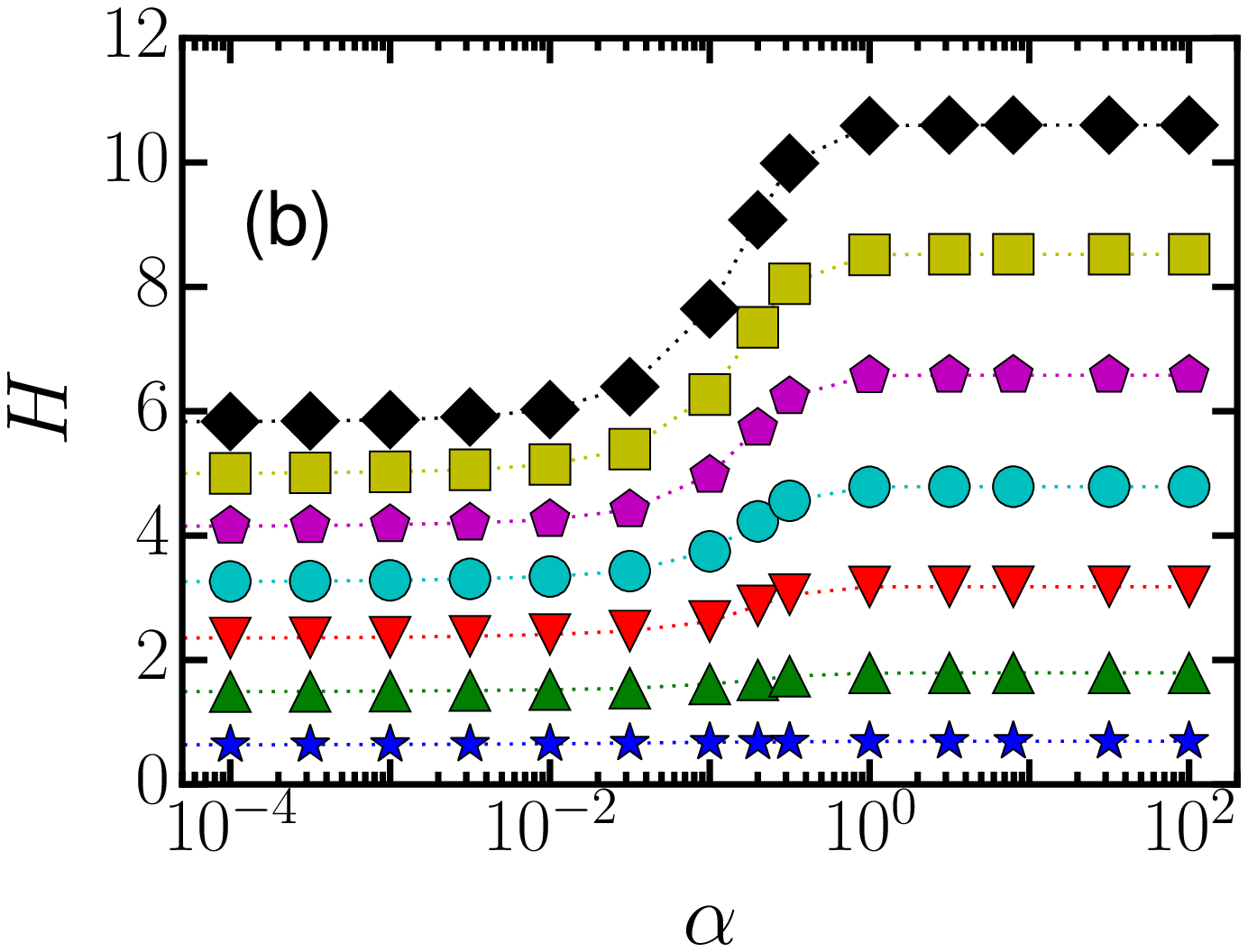}
  \caption{Permutation entropy ($H$) as a function of the size of the 
	ordinal pattern ($D$) and the noise strength ($\alpha$) for data 
	generated from the Logistic map. (a) $H$ vs $D$ for $\alpha= 
	1\times10^{-4}$ (stars), $\alpha= 2\times10^{-2}$ (triangles), 
	$\alpha= 5\times10^{-2}$ (inverted triangles), $\alpha=0.1$ 
	(circles), $\alpha=$1 (pentagons) and $H_{max}=\ln D!$ (solid 
	line). (b) $H$ versus $\alpha$ for $D=2$ (stars), $D=3$ 
	(triangles),  $D=4$ (inverted triangles), $D=5$ (circles), $D=6$ 
	(pentagons), $D=7$ (squares) and $D=8$ (diamonds).}
 \label{fig1}
\end{figure}

Figure \ref {fig1}a displays the permutation entropy, $H$, vs the 
dimension of the ordinal patterns, $D$, computed from time series of the 
logistic map at different noise strengths. It can be observed that $H$ 
increases monotonically with $D$, regardless of the noise strength. As 
the noise increases, $H$ approaches its maximum value, corresponding to 
equally probable ordinal patterns, $H_{max}=\ln D!$ (solid black line). 
Note that at ${\alpha = 1}$ (pentagons) the values of $H$ is already 
very close to $H_{max}$. Figure \ref {fig1}b displays $H$ as a function 
of the noise strength $\alpha$. A clear transition from low-noise to 
high-noise can be observed, for a value of the noise strength 
approximately independent of $D$. The difference between the values of 
the entropies at low and high noise becomes more pronounced as $D$ 
increases.

\begin{figure}[tp]
\centering
  \includegraphics[width=0.8\textwidth]{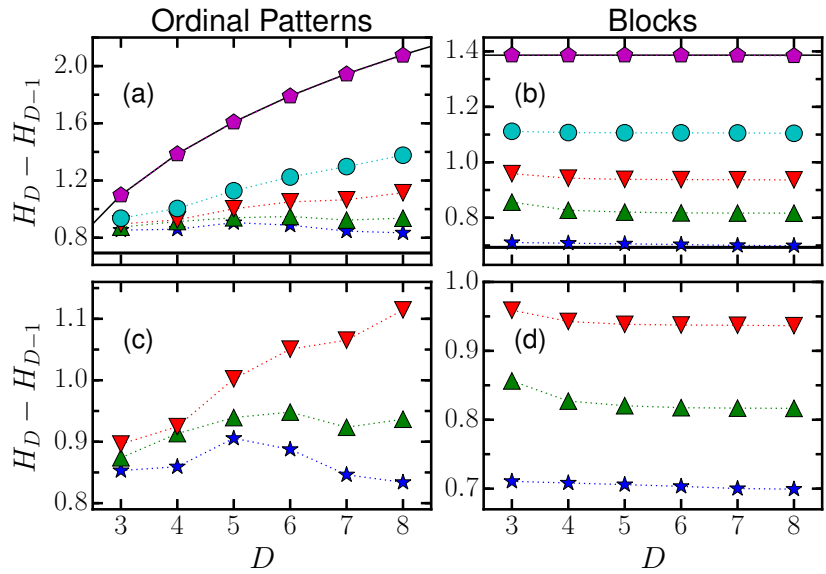} 
  \caption{Comparison of the entropy computed from ordinal patterns, and 
	the entropy computed from the blocks, for the Logistic map. The 
	difference $H_D-H_{D-1}$ is plotted vs. the dimension of the 
	ordinal patterns (a,c) and of the blocks (b,d) for various 
	values of noise strength [the noise strengths are as in Fig. 
	(\ref{fig1}a)]. In panels (a) and (b) the solid lines indicate 
	the asymptotic values for low noise (thick) and high noise 
	(thin). Panel (c) and (d) display a detail of (a) and (b).}
 \label{fig2}
\end{figure}

To further investigate this transition, Fig. (\ref {fig2}a) displays the 
difference $H_D-H_{D-1}$ as a function of $D$, for various values of 
noise strength. As before, we indicate with a thin black line the 
noise-dominated limit in which all patterns are equiprobable, $H_D - 
H_{D-1} = (\ln D! - \ln (D-1)!)$. In the opposite limit of 
almost-deterministic, as $D$ grows the expected value of $H_D-H_{D-1}$ is 
the Kolmogorov-Sinai entropy \cite{KBP_2002}, which for a one-dimensional 
chaotic map is equal to the Lyapunov exponent $\lambda$. In the case of 
logistic map for a local parameter set at $4$ one has $\lambda =\ln2$, 
indicated by the thick black line. As shown in detail in Fig. 
(\ref{fig2}c), we identify three possibilities:
\begin{itemize}
  \item a almost-deterministic regime in which $H_D-H_{D-1}$ decreases 
  for large $D$,
  \item a noise-dominated regime in which $H_D-H_{D-1}$ increases for 
  large $D$,
  \item an intermediate regime in which $H_D-H_{D-1}$ remains nearly 
  constant with $D$.
\end{itemize}

In principle, this qualitative feature of the permutation entropy can be 
applied to experimental time series to assess whether the dynamics is 
dominated by noise or by the deterministic dynamics.

We remind that this distinction can not be done for the block entropy, as 
in this case $H_D-H_{D-1}$ is necessarily a decreasing function of $D$ 
(see e.g. \cite{shannon48,1056823}). This fundamental difference between 
permutation entropy and block entropy can be appreciated by comparing 
the left and right panels of Fig. (\ref{fig2}).

\begin{figure}[tp]
  \centering
  \includegraphics[width=0.495\textwidth]{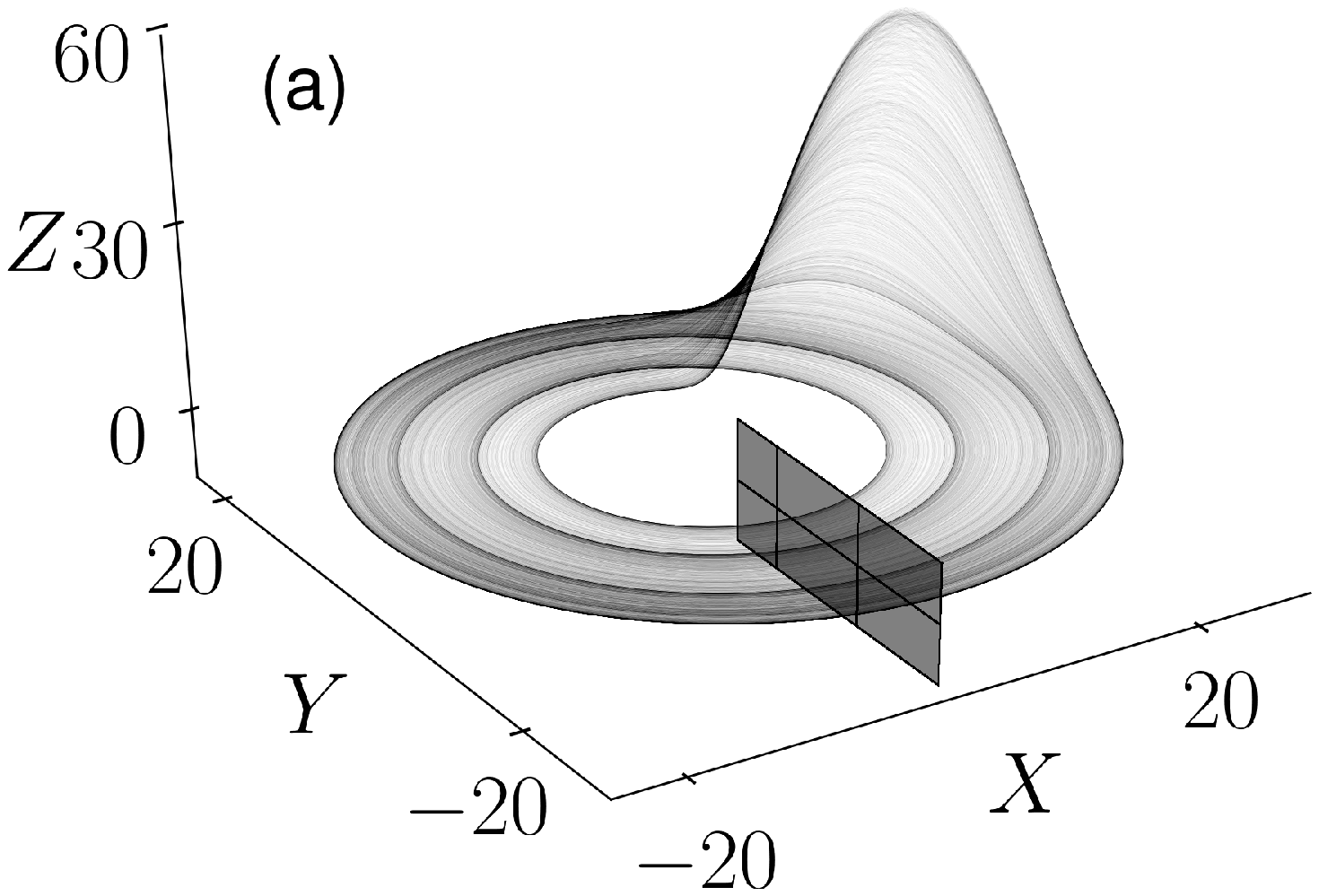}
  \includegraphics[width=0.495\textwidth]{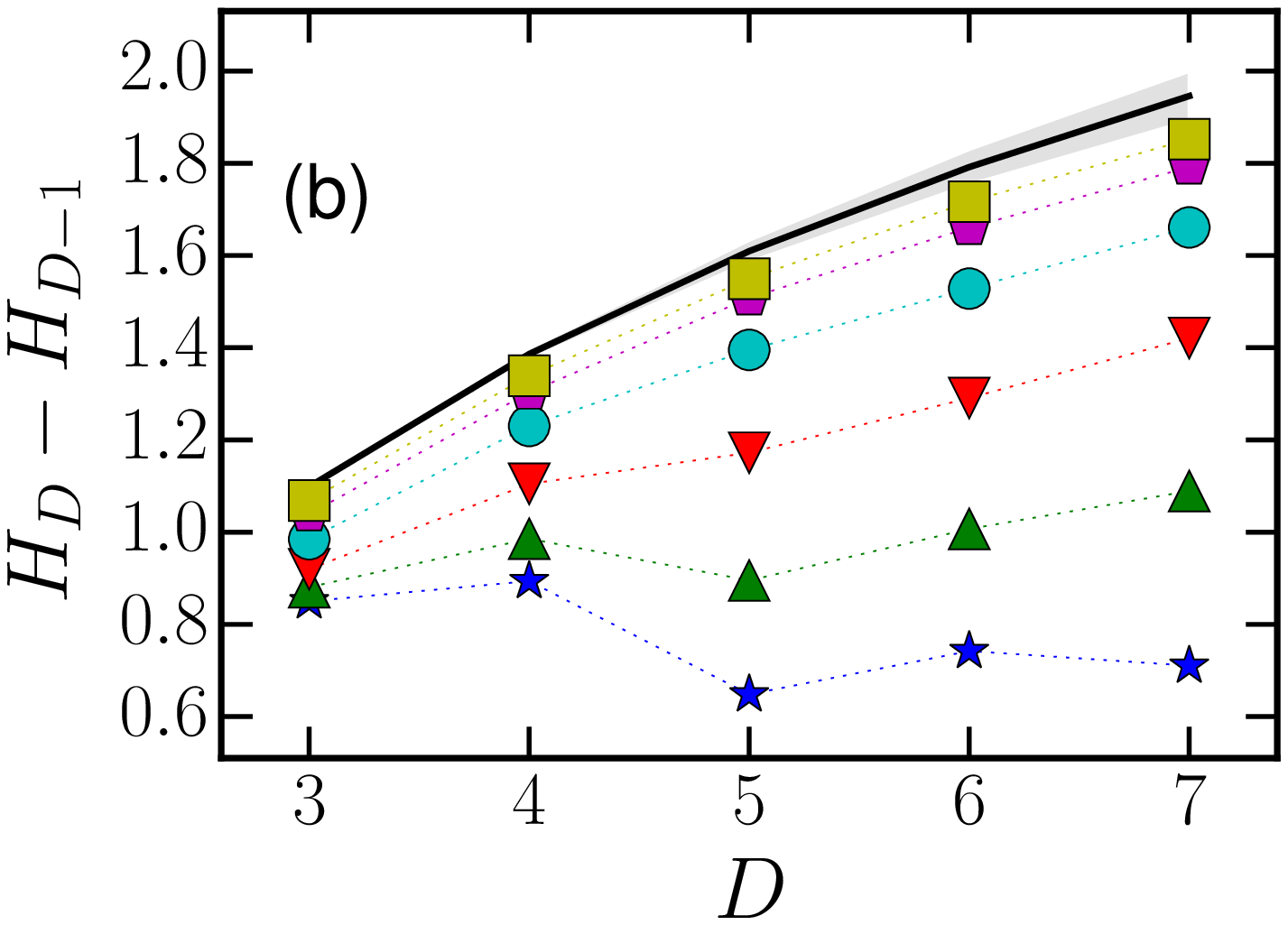}
  \caption{(a) R\"ossler attractor and Poincar\'e section in $X =  0$. 
      (b) Permutation entropy difference, $H_D-H_{D-1}$, vs the 
      dimension of the ordinal patterns, $D$, for noise strength $\alpha 
      = [0$ (star), $0.8 $ (triangle), $ 1.6 $ (inverted triangle), 
      $2.4$ (circle), $ 3.2 $ (pentagon), $ 4$ (square)$]$. The gray 
      region indicates the values of $H_D-H_{D-1}$ that are consistent 
      with equally probable ordinal patterns (see text for details). For 
      the smallest value of alpha, $H_D-H_{D-1}$ shows a non-monotonic 
      behavior, while  for higher values of the noise strength, 
      $H_D-H_{D-1}$ grows monotonically with $D$.}
  \label{fig3}
\end{figure}

For the analysis of the Rossler data, we considered the Poincar\'e map 
$X=0$, shown in Fig. (\ref {fig3}a), and analyzed the sequence of 
time-intervals between consecutive crossings. Figure (\ref{fig3}b) shows 
the difference $H_D-H_{D-1}$ vs $D$, for different values of $\alpha$. 
The solid line indicates the expected value if all ordinal patterns were 
equally probable, $H_D-H_{D-1} =\ln D!-\ln (D-1)!$. Because of the high 
level of stochasticity, we calculate the confidence interval that is 
consistent with the null hypothesis of equally probable ordinal 
patterns: in Fig. (\ref{fig3}b) the gray region represents the expected 
value $\pm 3 \sigma$, where $\sigma$ is the standard deviation calculated 
for a hundred surrogated (shuffle) time-series.

\begin{figure}[tp]
  \centering
  \includegraphics[width=0.495\textwidth]{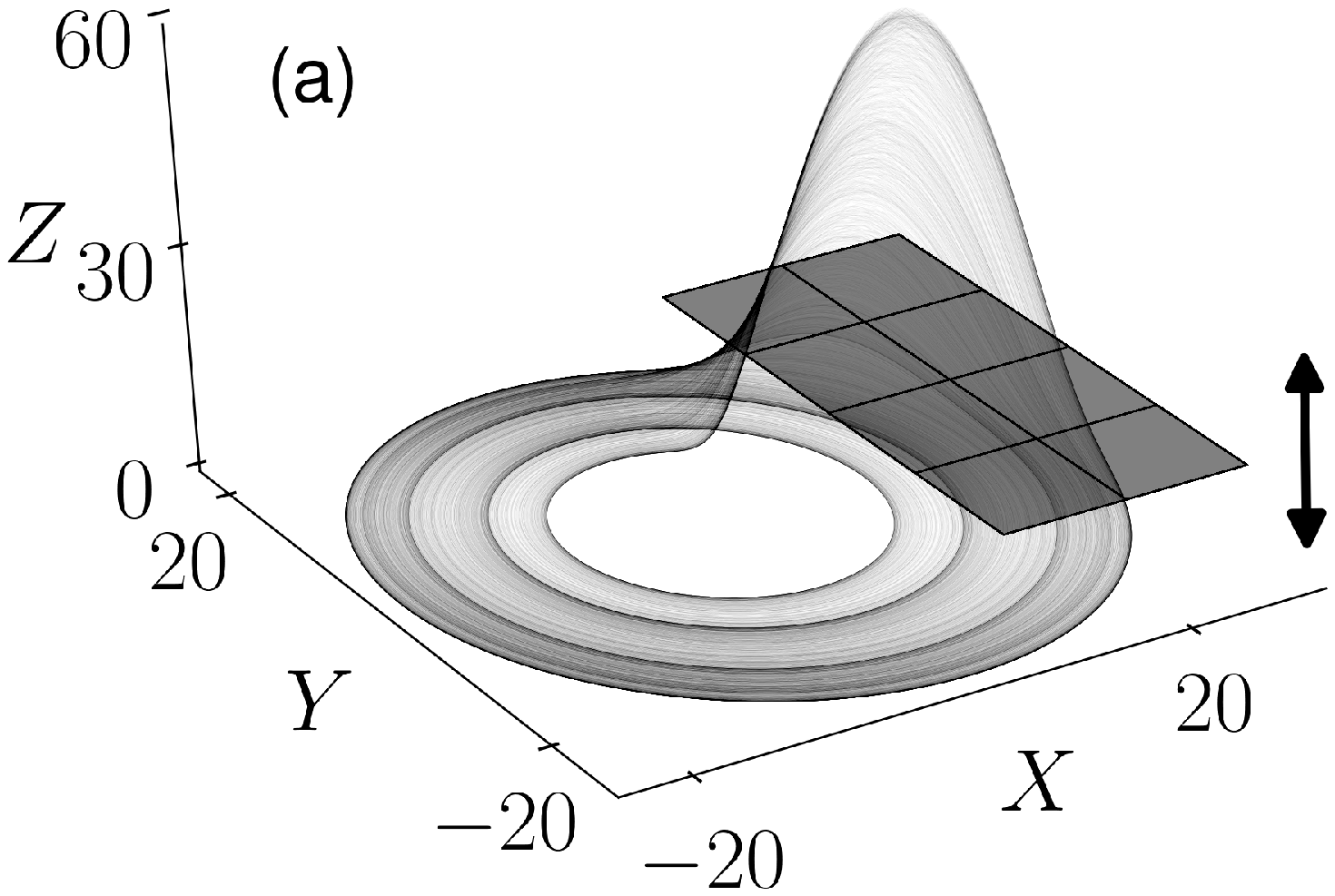}
  \includegraphics[width=0.495\textwidth]{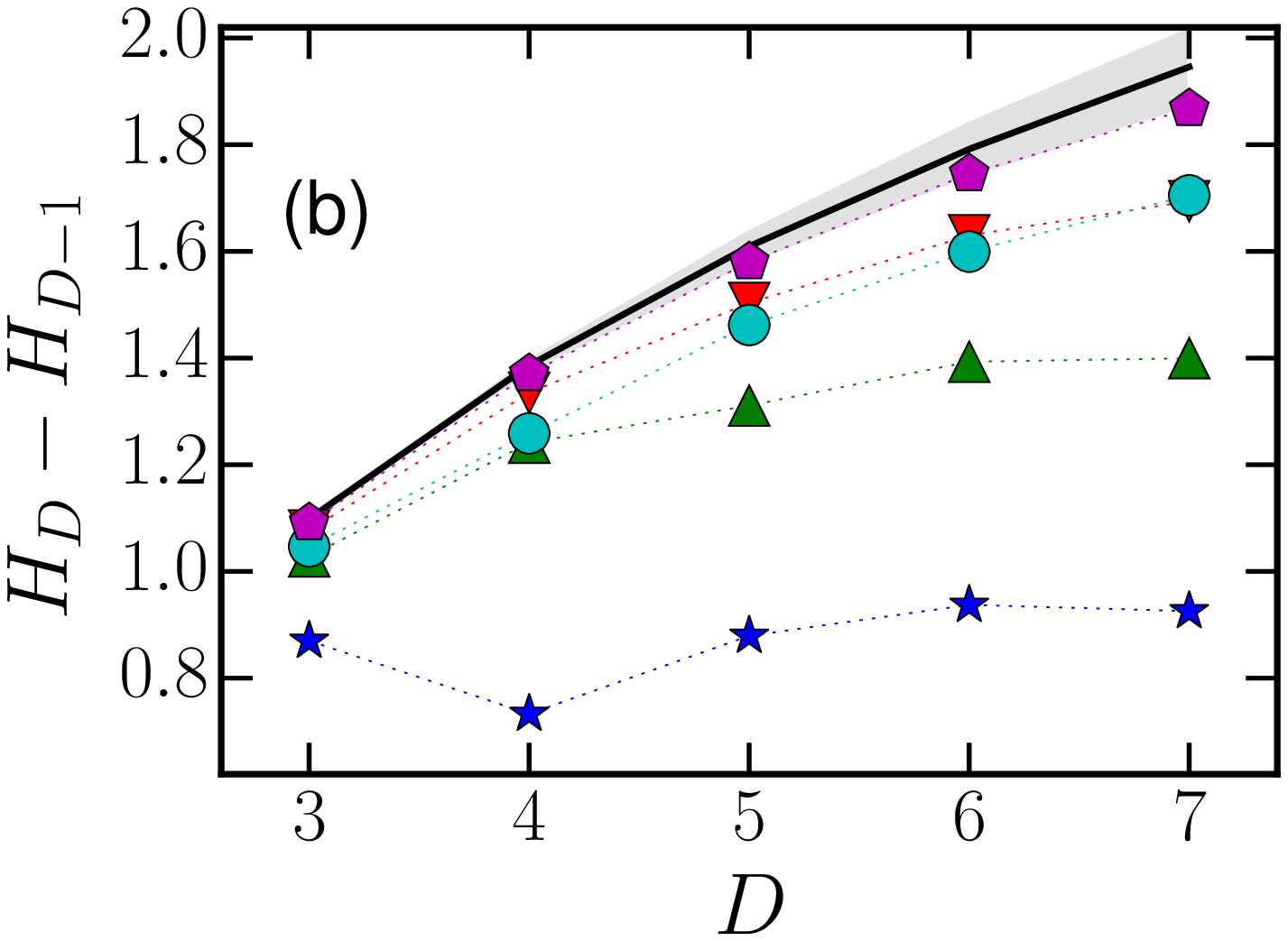}
  \caption{(a) R\"ossler attractor and Poincar\'e section placed in $ z 
	= \beta $. (b) Permutation entropy difference, $H_D-H_{D-1}$, 
	vs the dimension of the ordinal patterns, $D$, for $\beta=0.05$ 
	(stars), $\beta= 6.7$ (triangles),   $\beta=13.4$ (inverted 
	triangles), $\beta= 20.0$(circles), $\beta= 26.7$ (pentagons). 
	The behavior is qualitatively similar to the one observed in 
	Fig. (\ref{fig3}b).}
  \label{fig4}
\end{figure}

Before testing the method in experimental data we want to investigate how 
the choice of the Poincar\'e section influences the results. We consider 
a Poincar\'e section in the plane $ Z = \beta $, as shown in Fig. 
(\ref{fig4}a), and varying $\beta$ in the range $[0.05-26.7]$, for a 
fixed value of $\alpha = 0$. In this case, to discretize the time series, 
we analyze the time values when the trajectory intersects the Poincar\'e 
section and $Z$ grows.

Figure (\ref {fig4}b) displays the difference $H_D-H_{D-1}$ vs. $D$, for 
different values of $\beta$. We can see that the difference $H_D - 
H_{D-1}$ increases with $\beta$. This is due to the fact that, as $\beta$ 
is increased, consecutive values in the time-series become increasingly 
uncorrelated, similarly to when increasing the noise strength. On the 
contrary, for the minimum value of $\beta$, the variation of $H_D - 
H_{D-1}$ with $D$ is resemblant to the behavior under 
almost-deterministic observed in Fig. (\ref {fig3}b).

\begin{figure}[tp]
  \centering
  \includegraphics[width=0.495\textwidth]{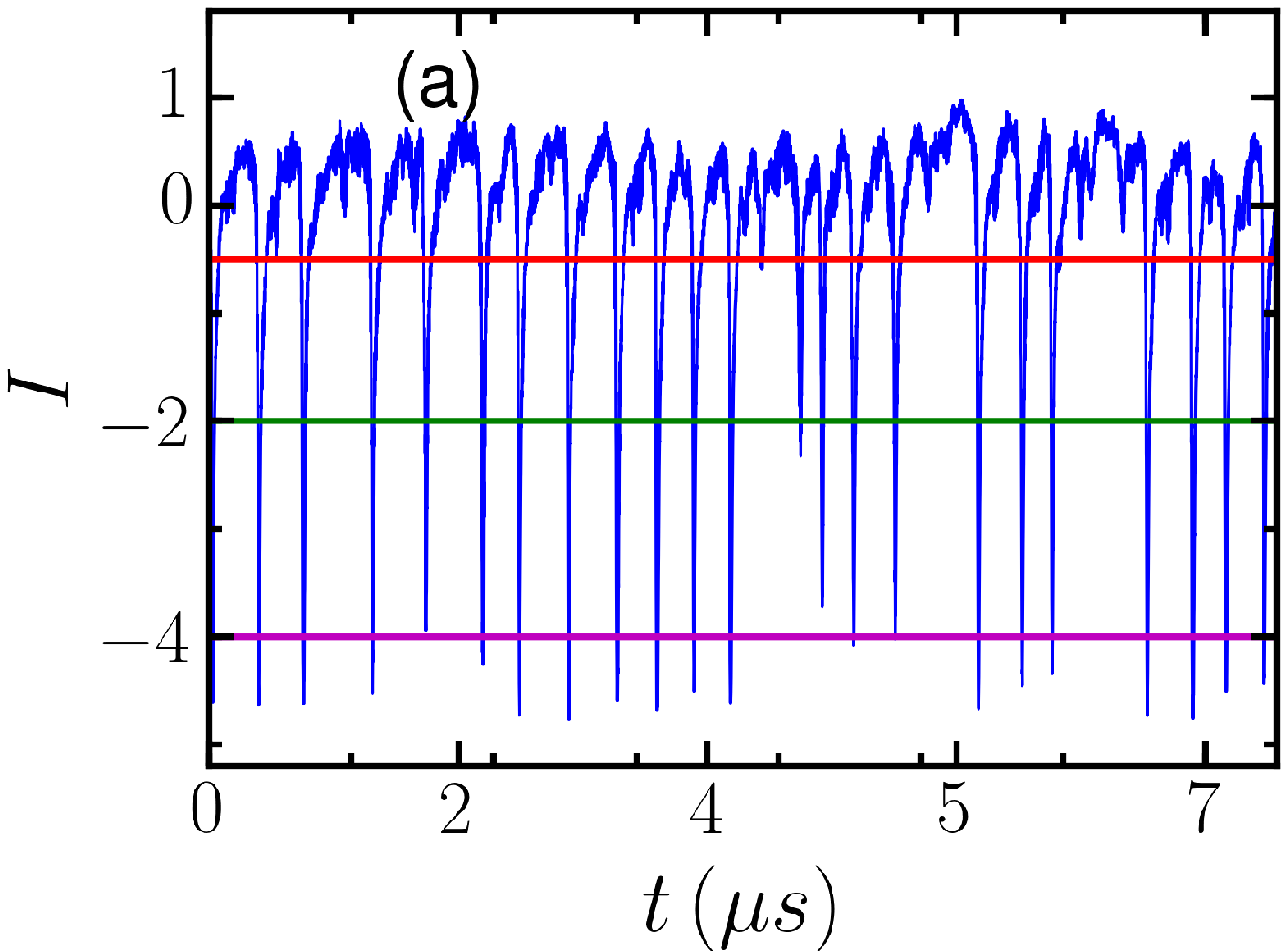}
  \includegraphics[width=0.495\textwidth]{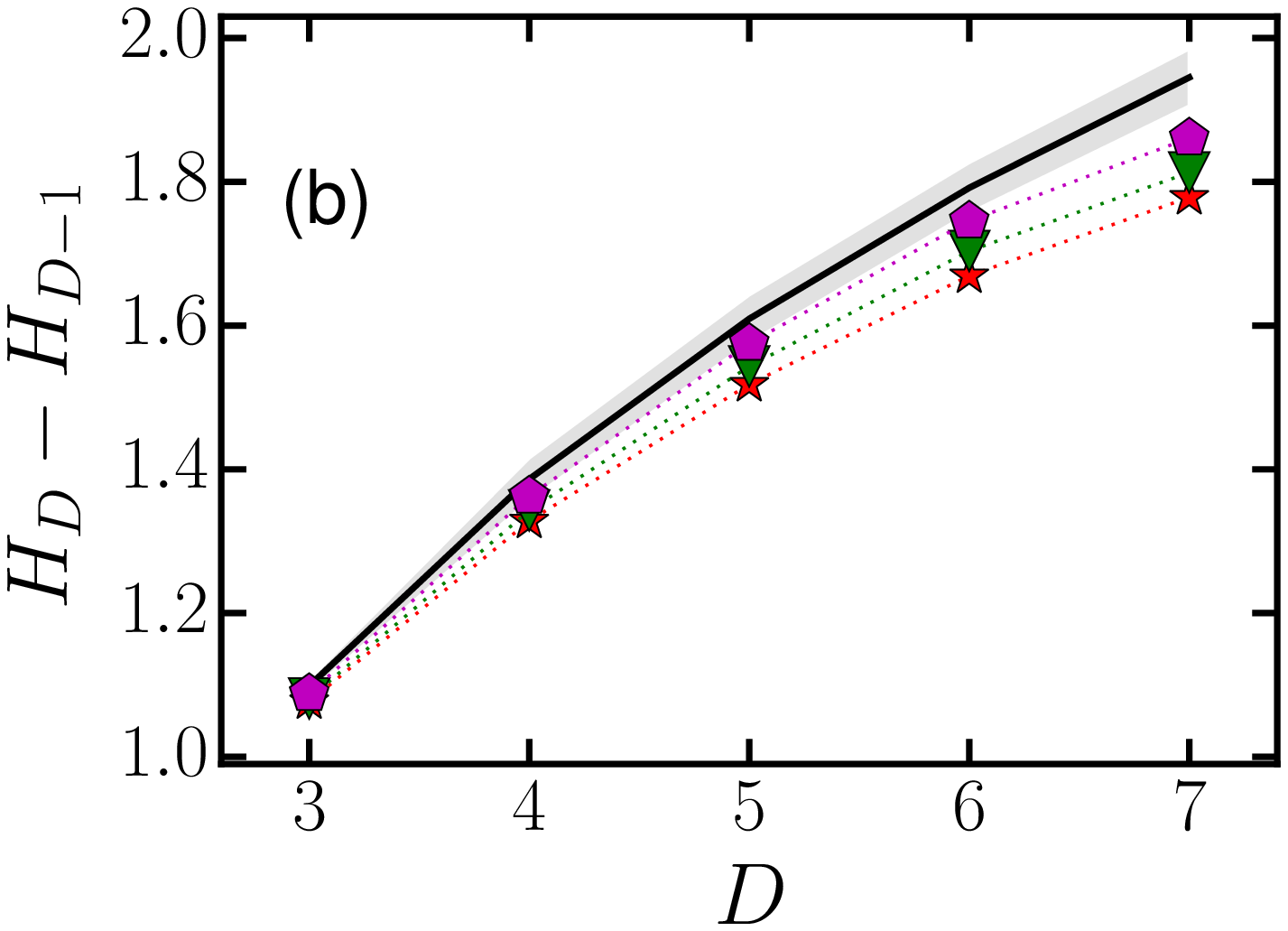}
  \caption{(a) Experimentally recorded time-series for the output 
      intensity of a semiconductor laser, which operates in the 
      low-frequency fluctuations (LFFs) regime, induced by self 
      time-delayed optical feedback. The horizontal lines indicate the 
      thresholds used to detect the dropout times. (b) Permutation 
      entropy difference, $H_D-H_{D-1}$, vs the dimension of the ordinal 
      patterns, for different thresholds: $-0.5$ (stars), $-2$ 
      (inverted triangles) and $-4$ (pentagons).}
  \label{fig5}
\end{figure}

Finally, we analyze experimental data from the laser output intensity, 
displayed in Fig. (\ref{fig5}a). To discretize the data we consider the 
thresholds indicated with horizontal lines in Fig. (\ref{fig5}a), and 
analyzed the time intervals between consecutive threshold-crossings 
\cite{tiana_pra_2010, aragoneses_2013, taciano_optics_express_2015}. 
Figure (\ref{fig5}b) displays the difference $H_D-H_{D-1}$ vs. $D$, for 
different thresholds. Note that $H_D-H_{D-1}$ varies with the threshold 
in a similar way as in Fig. (\ref{fig4}b): as the threshold decreases, 
correlations between consecutive dropouts are lost.

For all the thresholds, $H_D-H_{D-1}$ grows monotonically with $D$. The 
reason is that the empirical time series is very noisy and the 
``almost-deterministic'' regime is not seen, not even for the highest 
threshold. Nevertheless, the values of $H_D-H_{D-1}$ lie outside the gray 
region that indicates values consistent with equally probable ordinal 
patterns. This reveals that the sequence of intensity dropouts are not 
completely uncorrelated, and thus, this method can determine regularities 
also in very noisy data.

\section{Conclusion}

We have studied the influence of noise in the permutation entropy of 
dynamical systems, considering both, simulated data and experimental 
data. In the simulated data, when increasing the noise strength, a 
transition between a almost-deterministic regime and a noise-dominated 
regime was clearly observed. The noise value at which this transition 
occurs is roughly independent of the size $D$ of the ordinal pattern.

In the almost-deterministic regime, the permutation entropy grows almost 
linearly or sub-linearly with $D$. This behavior is qualitatively similar 
to that of the block entropy. However, to observe a quantitative 
equivalence it is often needed to analyze extremely long time series, 
which can be computationally unfeasible even for relatively simple 
dynamical systems. In the noise-dominated regime, the growth is faster 
than linear, i.e. the differences $H_D-H_{D-1}$ increase with $D$. In 
principle, this fact can be used to determine whether the dynamics is in 
a noise-dominated or a almost-deterministic regime from an experimental 
time series where the noise strength can not be externally tuned. 
However, care must be taken in interpreting the results, as extracting a 
one-dimensional time series from a purely deterministic high-dimensional 
time series via a Poincar\'e map can lead to ordinal patterns which look 
effectively noisy, as we demonstrated in the example of the R\"ossler 
systems. This fact reflects the well-known difficulties of distinguishing 
deterministic dynamics from noise when dealing with high-dimensional 
systems \cite{cencini_2000}.

\section{Acknowledge}
This work was supported by grants EOARD FA9550-14-1-0359, Ministerio de 
Ciencia e Innovacion, Spain and FEDER, FIS2012-37655-C02-01 and the 
Marie Curie Initial Training Network NETT, FP7-PEOPLE-2011-ITN 289146.


\end{document}